\documentclass[12pt]{article}

\usepackage{amsthm,amsmath,amssymb}

\usepackage{graphicx}
\usepackage[multiple]{footmisc}

\usepackage[colorlinks=true,citecolor=black,linkcolor=black,urlcolor=blue]{hyperref}

\theoremstyle{plain}
\newtheorem{theorem}{Theorem}[section]

\newtheorem{proposition}[theorem]{Proposition}

\theoremstyle{definition}
\newtheorem{definition}[theorem]{Definition}
\newtheorem{example}[theorem]{Example}

\theoremstyle{remark}
\newtheorem{remark}[theorem]{Remark}

\date{}

\title{\bf A Brief Introduction to Shannon's Information Theory}

\author{Ricky Xiaofeng Chen\footnote{School of Mathematics, Hefei University of Technology, Hefei, Anhui, P.R. China.
	{\it Email}:	chen.ricky1982@gmail.com, chenshu731@sina.com} \footnote{The author was a wireless research engineer working for Huawei during 2008--2013.}
\\
}

\begin{document}

\maketitle

\begin{abstract}

This article serves as a brief introduction to the Shannon information theory.
Concepts of information, Shannon entropy and channel capacity are mainly covered. All these concepts are developed in a totally combinatorial flavor.
Some issues usually not addressed in the literature are discussed here as well.
In particular, we show that it seems we can define channel capacity differently which allows us to potentially transmit
more messages in a fixed sufficient long time duration. However, for a channel carrying a finite number of letters, the channel capacity unfortunately remains the same as the Shannon limit.


\end{abstract}

\section{Preface}
Claud Shannon's paper ``A mathematical theory of communication"~\cite{shannon1}
published in July and October of 1948 is
the Magna Carta of the information age. Shannon's discovery
of the fundamental laws of data compression and transmission
marks the birth of Information Theory.

In this note, we first discuss how to formulate the main fundamental quantities in Information Theory: information, Shannon entropy and channel capacity.
 We then present the derivation of the classical capacity formula under the channel with additive white Gaussian noise (AWGN).
 For more relevant detailed introduction, we refer the readers to the material~\cite{cover-thomas,shannon1,shannon2,tse,vd,witten} and references therein.
 Meanwhile,
we have some discussion concerning whether the Shannon limit can be broken or not.
In particular, we argue that a new way of defining channel capacity taking account of all admissible sets of distributions seems reasonable, and we show that the corresponding channel capacity is unfortunately the same as the Shannon limit for channels carrying a finite number of symbols.

\section{Information and Entropy}
What is information? or,
 what does it mean when Michael says he has gotten some information regarding something?

  Well, it means that he did not know what this ``something" is about before someone else ``communicate" some stuff about this ``something" to him. But now, after the communication, he knows it. Note that anything can be described by several sentences in a language, for instance, English. A sentence or sentences in English can be viewed as a sequence of letters (i.e., `a',`b',`c',\ldots) and symbols (i.e., `,',`.',`\textvisiblespace',\ldots). Thus, in theory, we can just think of sentences conveying different meaning as different sequences (of letters and symbols).

``Michael is not sure of what this ``something" is about" can be understood as ``Michael is not sure to which sequence this ``something" corresponds". Of course, we can assume that he is aware of all possible sequences, only which one of them remains uncertain w.r.t. this ``something". Michael can get some information when someone else ``pick" one sequence (the one conveying the meaning someone else has in mind) out of all possible sequences and ``communicate" it to Michael.
In this sense, we can say that this sequence, even each letter there, contains certain amount of information.

Another aspect of these sequences is that not all sequences, words, or letters appear equally. They appear following some probability distribution. For example, the sequence ``how are you" is more likely to appear than ``ahaojiaping mei"; the letter `e' is more likely to appear than the letter `z' (the reader may have noticed that this is the first time the letter `z' appears in the text so far).

The rough ideas above are the underlying motivation of the following more formal discussion on what information is, how to measure information, and so on.

\subsection{How many sequences are there}

In order to formalize the ideas we have just discussed, we assume there is an alphabet $\mathbb{A}$ of $n$ letters, i.e., $\mathbb{A}=\{x_1,x_2,\ldots , x_n\}$.
For example, $\mathbb{A}=\{a,b,\ldots,z, \text{`,'},\text{`.'},\text{`\textvisiblespace'},\ldots\}$, or just as simple as $\mathbb{A}=\{0,1\}$.
We will next be interested in a set of sequences with entries from the alphabet.
We assume the letter $x_i$ ($1\leq i \leq n$) appears in each position of the sequence interested with probability $0\leq p_i\leq 1$, and $\sum_{i=1}^n p_i=1$.
To make it simple, we further assume that for any such sequence $s=s_1s_2s_3\cdots s_T$, where $s_i=x_j$ for some $j$, the exact letters taken by different entries $s_i$ and $s_j$ are independent for all $i\neq j$.

{\noindent {\bf Now comes to the fundamental question:} with these assumptions, how many possible desired sequences are there?

It should be noted that a short sequence (i.e.,~$T$ is small) consisting of these letters from the alphabet $\mathbb{A}$ will not properly and explicitly reflect the statistical properties we have assumed above. Thus, the length $T$ of these sequences that we are interested should be quite large, and we will consider the situation as $T$ goes to infinity, denoted by $T\rightarrow \infty$. From the viewpoint of statistics, each sequence of length $T$ can be equivalently viewed as a series of $T$ independent experiments and the possible outcomes of each experiment are these events (i.e.,~letters) in $\mathbb{A}$, where the event $x_i$ happens with probability $p_i$. By the Law of Large Numbers, for $T$ large enough, in each series of $T$ independent experiments, the event $x_i$ will (almost surely) appear $T\times p_i$ ($Tp_i$ for short) times. Assume we label these experiments by $1,2,\ldots ,T$. Now, \emph{the only thing we do not know is in which experiments the event $x_i$ happens.}

Therefore, the number of sequences we are interested is equal to the number of different ways of placing $Tp_1$ number of $x_1$, $Tp_2$ number of $x_2$, and so on, into $T$ positions such that each position contains only one letter.
Equivalently, it is the number of different ways of placing $T$ different balls into $n$ different boxes such that there are $Tp_1$ balls in the first box, $Tp_2$ balls in the second box, and so on and so forth.

Now it should be easy to enumerate these sequences. Let's first consider a toy example:

\begin{example}
	Assume there are $T=5$ balls and $2$ boxes. How many different ways to place $2$ balls in the first box and $3$ balls in the second? The answer is that there are in total ${5\choose 2}{5-2 \choose 3}=10$ different ways, where
	${n\choose m}=\frac{n!}{m!(n-m)!}$ and $n!=n\times (n-1) \times \cdots  \times 1$.
\end{example}

The same as the above example, for our general setting here, the total number of sequences
we are interested is
\begin{align}
K={T\choose Tp_1}\times {T-Tp_1\choose Tp_2}\times {T-Tp_1-Tp_2 \choose Tp_3}\times \cdots \times
{T-Tp_1-\cdots -Tp_{n-1} \choose Tp_n}.
\end{align}

\subsection{Average amount of required resource for storage}
Next, if we want to uniquely index each sequence among these $K$ sequences using binary digits, i.e., a sequence using only $0$ and $1$, what is the minimum length of the binary sequence?
Let us still look at an example first.
\begin{example}
	If $K=4$, all $4$ sequence can be respectively indexed by $00$,~$01$,~$10$ and $11$. Certainly, we cannot use only one digit to give a unique index for each and every sequence.
	So, the binary sequence should have a length at least $\log_2 4=2$.
\end{example}
Therefore, the binary sequence should have a length $\log_2 K$ in order to uniquely index each and every sequence among all these $K$
sequences. In terms of Computer Science, we need $\log_2 K$ bits to index (and store) a
sequence. Next, we will derive a more explicit expression of $\log_2 K$.

If $m$ is large enough, $m!$ can be quite accurately approximated by the {\it Stirling formula}:
\begin{align}
	m!\approx \sqrt{2\pi m}\left(\frac{m}{e}\right)^m.
\end{align}
For fixed $a,b \geq 0$, it is true that $Ta, Tb\rightarrow \infty$ as $T\rightarrow \infty$. Then we have the approximation:
\begin{align}
{Ta\choose Tb}=\frac{(Ta)!}{(Tb)!(Ta-Tb)!}& \approx \frac{\sqrt{2\pi Ta}\left(\frac{Ta}{e}\right)^{Ta}}{\sqrt{2\pi Tb}\left(\frac{Tb}{e}\right)^{Tb}\sqrt{2\pi T(a-b)}\left(\frac{T(a-b)}{e}\right)^{T(a-b)}}\nonumber\\
&=\frac{\sqrt{a}a^{Ta}}{\sqrt{2\pi T}\sqrt{b(a-b)}b^{Tb}(a-b)^{T(a-b)}},
\end{align}
and
\begin{align}
\log_2 {Ta\choose Tb} \approx -\log_2 \sqrt{2\pi T} + \log_2 \sqrt{a} -\log_2 \sqrt{b} -\log_2 \sqrt{a-b} \nonumber\\+Ta\log_2 a -Tb \log_2 b -T(a-b) \log_2 (a-b).
\end{align}
Notice that for any fixed $p_i> 0$, $Tp_i\rightarrow \infty$ as $T\rightarrow \infty$, which means we can apply the approximation eq.~$(4)$ to every factor in eq.~$(1)$. By doing this, we obtain
\begin{align}
\log_2 K\approx -n\log_2 \sqrt{2\pi T}-\log_2 \sqrt{p_1}-\log_2 \sqrt{p_2}-\cdots -\log_2 \sqrt{p_n}\nonumber\\
-Tp_1 \log_2 p_1-Tp_2 \log_2 p_2-\cdots -Tp_n \log_2 p_n.
\end{align}
Next if we consider the average number of bits that a letter needs in indexing a sequence of length $T$, {\bf a minor miracle happens}: as $T\rightarrow \infty$,
\begin{align}\label{entropy}
\frac{\log_2 K}{T}\approx -\sum_{i=1}^n p_i \log_2 p_i.
\end{align}
The surprisingly simple expression on the right hand side (RHS) of eq.~\eqref{entropy} is the celebrated quantity associated with a probability distribution, called {\emph{Shannon entropy}}.

Let us review a little bit what we have done. We first have $K$ sequences interested in total, and all sequences appear equally likely. Suppose they encode different messages. Regardless of the specific messages they encode, we view them as having the same amount of information. Then it is natural to employ the number of bits needed to encode a sequence to measure the amount of information a sequence encode (or can provide). Then, the quantity $\frac{\log_2 K}{T}$ can be viewed as the average amount of information a letter in the sequence has.
This suggests that we can actually define the amount of information of each letter.
Here, we say ``average'' because we think the amount of information that different letters have should be different as they may not ``contribute equally" in a sequence, depending on the respective probabilities of the letters. Indeed, if we look into the RHS of the formula~\eqref{entropy}, it only depends on the probability distribution of these letters in $\mathbb{A}$. Note
$$
-\sum_{i=1}^n p_i \log_2 p_i=\sum_{i=1}^n p_i \times {\log_2 \frac{1}{p_i}}
$$
is clearly the expectation (i.e., average in the sense of probability) of the quantity ${\log_2 \frac{1}{p_i}}$
associated with the letter $x_i$, for $1\leq i \leq n$. This matches the term ``average" so that we can define
\emph{the amount of information} that a letter $x_i$ appearing with probability $p_i$ has to be ${\log_2 \frac{1}{p_i}}$ bits.

In this definition of information, we observe that if a letter has a higher probability it has less information, and vice versa.
In other words, more uncertainty, more information.
Just like lottery, winning the first prize is less likely but more shocking when it happens, while you may feel winning a prize of 10 bucks is not a big deal since it is very likely. Hence, this definition agrees with our intuition as well.

In the subsequent of the paper, we will omit the base in the logarithm function.
Theoretically, the base could be any number and is $2$ by default.
Now we summarize information and Shannon entropy in the following definition:

\begin{definition}
	Let $X$ be a random variable, taking value $x_i$ with probability $p_i$, for $1\leq i \leq n$.
	Then, the quantity $I(p_i)={\log \frac{1}{p_i}}$ is the amount of {\emph{information}} encoded in $x_i$ (or $p_i$),
	while the average amount of information $\sum_{i=1}^n p_i \times {\log \frac{1}{p_i}}$ is called the {\emph{Shannon entropy}}
	of the random variable $X$ (or the distribution $P$), and denoted by $H(X)$.
\end{definition}

To the best of our knowledge, the approach of obtaining the expression for the Shannon entropy we presented above seems having not been discussed much in the literature. Usually, the expression for information is first derived via certain approach, and then the expression for entropy follows as the expectation.

Note that from the derivation of the Shannon entropy, if a distribution $X$ has Shannon entropy $H(X)$,
then there are approximately $K=2^{T[H(X)+o(1)]}$ sequences satisfying the distribution for $T$ being large enough.

\noindent\emph{\bf Question}: among all possible probability distributions on at most $n$ letters, which distributions give the largest Shannon entropy?
For finite case, the answer is given in the following proposition.
\begin{proposition}\label{fntmax}
	For finite $n$, when $p_i=\frac{1}{n}$ for $1\leq i \leq n$, the Shannon entropy attains the
	maximum
$$
\sum_{i=1}^n \frac{1}{n} \times {\log n}=\log n.
$$
\end{proposition}

\subsection{Further definitions and properties}
The definition of information and entropy can be extended to continuous random variables.
Let $X$ be a random variable taking real (i.e., real numbers) values and let $f(x)$ be its probability density function.
Then, the probability $P(X=x)=f(x)\Delta x$ with $\Delta x$ being very very small. Mimic the discrete finite case, the entropy of $X$
can be defined by
\begin{align}
H(X)=\sum_x -P(x)\log P(x) &=\lim_{\Delta x\rightarrow 0} \sum_x -
[f(x)\Delta x] \log [f(x)\Delta x]\\
&=\lim_{\Delta x\rightarrow 0} \sum_x -
[f(x)\Delta x] (\log f(x)+ \log \Delta x])\\
&= -\int f(x)\log f(x) \text{d}x -\log \text{d}x,
\end{align}
where we have used the definition of (Riemann) integral and the fact that $\int f(x)\text{d}x=1$.
The last formula~$(9)$  above is called the absolute entropy for the random variable $X$. Note that regardless of the probability distribution, there is always a positive infinity
term $-\log \text{d}x$. So, we can drop this term and define the (relative) entropy of $X$ to be
$$
-\int f(x)\log f(x) \text{d}x.
$$
When we discuss entropy for the continuous case, we usually use the relative entropy.
Next we may ask which continuous (infinite) distribution gives the maximum entropy?

\begin{proposition}\label{entropy-continuous}
	Among all real random variables with expectation $\mu$ and variance $\sigma^2$, the Gaussian distribution $X\sim \mathcal{N}(\mu, \sigma^2)$ attains the maximum entropy
$$
H(X)=-\int \frac{1}{\sqrt{2\pi \sigma^2}}e^{-\frac{(x-\mu)^2}{2\sigma^2}} \log \frac{1}{\sqrt{2\pi \sigma^2}}e^{-\frac{(x-\mu)^2}{2\sigma^2}} \text{d}x = \log \sqrt{2\pi \text{e} \sigma^2}.
$$
\end{proposition}

Note that joint distribution and conditional distribution are still just probability distributions.
Then, we can define entropy there accordingly.

\begin{definition}
	Let $X$ and $Y$  be two random variables with the joint distribution $P(X=x, Y=y)$ ($P(x,y)$ for short).
	Then the \emph{joint entropy} $H(X,Y)$ is defined by
	\begin{align}
	H(X,Y)=-\sum_{x,y} P(x,y) \log P(x,y).
	\end{align}
\end{definition}

\begin{definition}
	Let $X$ and $Y$ be two random variables with the joint distribution $P(x, y)$ and the conditional distribution
	$P(x\mid y)$.
	Then the \emph{conditional entropy} $H(X \mid Y)$ is defined by
	\begin{align}
	H(X \mid Y)=-\sum_{x,y} P(x,y) \log P(x \mid y).
	\end{align}
\end{definition}

\begin{remark}
	Fixing $X=x$, $P(Y \mid x)$ is also a probability distribution. It's entropy equals
	$$
	H(Y \mid x)= -\sum_y P(y \mid x) \log P(y \mid x)
	$$
	which can be viewed as a function over $X$ (or a random variable depending on $X$).
	It can be checked that $H(Y \mid X)$ is actually the expectation of $H(Y \mid x)$, i.e.,
	$$
	H(Y \mid X)= \text{E}_x \{H(Y\mid x)\}=\sum_x P(x) H(Y\mid x),
	$$
	using the fact that $P(x,y)=P(x)P(y \mid x)$.
\end{remark}

\begin{example}
If $Y=X$, we have
\begin{align*}
H(X\mid Y)=H(X\mid X)&=-\sum_{x,y} P(x,y) \log P(x \mid y)\\
&=- \sum_x P(x,x)\log P(x \mid x)=0,
\end{align*}
where we used the fact that
$$
P(x \mid y)=\left\{
\begin{array}{ll}
1  & \mbox{if } x = y, \\
0 & \mbox{if } x \neq y.
\end{array}
\right.
$$
This example shows, if a random variable $X$ is completely determined by another random variable $Y$, the uncertainty of $X$ after knowing $Y$ vanishes.

\end{example}

\begin{example}
If $Y$ and $X$ are independent, we have
\begin{align*}
H(X\mid Y)&=-\sum_{x,y} P(x,y) \log P(x \mid y)\\
&=- \sum_y\sum_{x} P(x)P(y)\log P(x)= H(X),
\end{align*}
where we used the fact that $P(x,y)=P(x)P(y)$ and $P(x \mid y)=P(x)$
for independent $X$ and $Y$.
This is the opposite case to the former example, saying if there is no connection between two random variables, the uncertainty of one remains unchanged even with the other completely known.
\end{example}

\section{Channel Capacity}

In a communication system, we have three basic ingredients: the source, the destination and the media between them. We call the media \emph{the (communication) channel}.
A channel could be in any form. It could be physical wires, cables, open environment in the case of wireless communication, antennas and certain combination of these.
In this section, we discuss channel capacity under channels without error and that with errors.

\subsection{Channel without error}
Given a channel and a set $\mathbb{A}$ of letters (or symbols) which can be transmitted via the channel, we suppose an information source generates letters in $\mathbb{A}$ following a probability distribution
$P$ (so we have a random variable $X$ taking values in $\mathbb{A}$), and send the generated letters to the destination through the channel.

Suppose the channel will carry the exact letters generated by the source to the destination. Then,
what is the amount of information received at the destination?
Certainly, the destination will receive exactly the same amount of information generated or provided by the source, which is $TH(X)$ in a time period of length of $T$ symbols (with $T$ large enough).
Namely, in a time period of symbol-length $T$, the source will generate
a sequence of length $T$, the destination will receive the same sequence, no matter what the sequence generated at the source is. Hence, the amount of information received at the destination is on average $H(X)$ per symbol.

The \emph{channel capacity} of a channel is the maximum amount of information on average can be obtained at the destination
in a fixed time duration, e.g., per second, or per symbol (time).
Put it differently, the channel capacity can be characterized by the maximum number of sequences on $\mathbb{A}$ that we can select
and transmit on the channel, such that
the destination can based on the received sequences, in principle, determine without error the corresponding sequences fed into the channel.

If the channel is errorless, what is the capacity of the channel? Well, as discussed above,
the maximum amount of information can be received at the destination equals the maximum amount of information can be generated at the source.
Therefore, the channel capacity $C$ for this case is
\begin{align}
C=\max_{X} H(X), \text{ per symbol},
\end{align}
where $X$ ranges over all possible distributions on $\mathbb{A}$.

For example, if $\mathbb{A}$ contains $n$ letters, then we know from
Proposition~\ref{fntmax} that the
uniform distribution achieves the channel capacity
$C=\log n$ bits per symbol.

\subsection{Channel with error}
What is the channel capacity of a channel with error?
A channel with error means that the source generated a letter $x_i \in \mathbb{A}$ and transmitted it to the destination via the channel, with some unpredictable error, the received letter at the destination may be $x_j$. Assume statistically, $x_j$ is received with probability $p(x_j \mid x_i)$ when $x_i$ is transmitted. These probabilities are called transit probabilities of the channel.
We assume that, once the channel is given, the transit probabilities are determined and will not change.

In order to understand the question better, we start with some examples.
\begin{example}\label{3ex1}
 Assume $\mathbb{A}=\{0,1\}$. If the transit probabilities of the channel are
 \begin{align*}
 p(1 \mid 0)=0.5, &\quad p(0 \mid 0)=0.5,\\
 p(1 \mid 1)=0.5, &\quad  p(0 \mid 1)=0.5,
 \end{align*}
 what is the channel capacity?

 The answer should be $0$, i.e., the destination cannot obtain any information at all.
 Because no matter what is being sent to the destination, the received sequence at the destination
 could be any $0-1$ sequence, with equal probability. From
 the received sequence, we can neither determine which sequence is the one generated at the source,
 nor can we determine which sequences are not the one generated at the source.

 In other words, the received sequence has no binding relation with the transmitted sequence on the channel at all,
 we can actually flip a fair coin to generate a sequence ourself instead of looking into
 the one actually received at the destination.
\end{example}

\begin{example}\label{3ex2}
	Assume $\mathbb{A}=\{0,1\}$. If the transit probabilities of the channel are
 \begin{align*}
 p(1 \mid 0)=0.1, &\quad p(0 \mid 0)=0.9,\\
 p(1 \mid 1)=0.9, &\quad  p(0 \mid 1)=0.1,
 \end{align*}
what is the channel capacity?
	The answer should not be $0$, i.e., the destination can determine something with regard to the transmitted sequence.
	
	Further suppose the source generates $0$ and $1$ with equal probability.
	Observe the outcome at the destination for a sufficient long time, that
	is a sequence long enough, for the illustration purpose, say a $10000$-letter long sequence is long enough (to guarantee
	the Law of Large Numbers to be effective).
	With these assumptions, there are approximately $5000$ $1$'s and $5000$ $0$'s, respectively, in the generated sequence at the source.
	Through the channel, $5000 \times 0.1=500$ $1$'s will change to $0$'s and vice versa.
	Thus, the received sequence should also have around $5000$ $1$'s and $5000$ $0$'s.

	Suppose the sequence received at the destination has $5000$ $1$'s for the first half of entries and $5000$ $0$'s for the second half of entries.
With these probabilities and received sequence known, what can we say about the generated sequence at the source? Well, it is not possible immediately to know what is the generated sequence based on these intelligences, because there are more than one sequence which can lead to the received sequence
	after going through the channel.
	But, the sequence generated at the source can certainly not be the sequence that contains
	$5000$ $0$'s for the first half and $5000$ $1$'s for the second half, or any sequence with most of $0$'s concentrating in the first half of entries.
	Since if that one is the generated one, the received sequence should contain about $4500$ $0$'s in the first half of entries in the received sequence, which is not the case observed in the received sequence.
	
	This is unlike Example~\ref{3ex1}, for which we can neither determine which is generated nor those not generated at the source. Thus, in the present example, the information obtained by the destination should not be $0$.

\end{example}

Let us come back to determine the capacity of the channel in general.
Recall the capacity is the maximum number of sequences on $\mathbb{A}$ that we can select and transmit on the channel such that
the destination can in principle determine without error the corresponding sequences fed into the channel based on the received sequences.
Since there is error in the transmission on the channel,
we can not select two sequences which potentially lead to the same sequence after going through the channel at the same time, otherwise we can never determine which one of the two is the transmitted one on the channel based on the same (received) sequence at the destination.

Hence, in order to determine the channel capacity, we need to determine the maximum number of sequences that are mutually disjoint, in the sense that any two will not lead to the same sequence at the destination.

Basically, the possible outputs at the destination are also sequences on $\mathbb{A}$, where element $x_i$, for $1\leq i \leq n$, appears in
these sequences with probability
$$p_Y(x_i)=\sum_{x_j\in \mathbb{A}} p(x_j) p(x_i \mid x_j) .
$$
Note this probability distribution will depend only on the distribution $X$ since
the transit probabilities are fixed.
Denote the random variable associating to this probability distribution at the destination by $Y(X)$ (note that $Y$ will change as $X$ change).

Shannon~\cite{shannon1} has proved that for a given distribution $X$, we can
choose at most
$$
2^{T[H(X)- H(X \mid Y)+o(1)]}
$$
 sequences (satisfying the given distribution) to be the sequences to transmit on the channel such that the destination can determine without error the transmitted sequence based on the received sequence.
That is, the destination can obtain $H(X)-H(X \mid Y)$ bits information per symbol.
The quantity $H(X)-H(X \mid Y)$ is called the \emph{mutual information} of $X$ and $Y$, denoted by $I(X,Y)$. It also holds that
$$
I(X,Y)=I(Y,X)=H(X)-H(X\mid Y)=H(Y)-H(Y \mid X).
$$
Therefore, the channel capacity for this case is
\begin{align}\label{capacity}
C=\max_X [H(X)- H(X \mid Y)], \text{ per symbol},
\end{align}
where $X$ ranges over all probability distributions on $\mathbb{A}$.
The quantity $C$ is called the \emph{Shannon capacity (limit)} of the channel (specified by the transit probability distribution).

Noticing that for a channel without error, $Y=X$ whence $H(X \mid Y)=0$ as discussed in Example~$2.9$, we realize the definition of capacity in eq.~\eqref{capacity} applies to channels without error as well.

\section{Capacity under AWGN}
In this section, we discuss the channel capacity of an AWGN channel of bandwidth $W$.
Here is how communication under an AWGN channel works: if $X=x$ is selected at the source and transmitted on the channel, at the destination side, $Y_x=x+W$ will be received, where $W$ is a Gaussian random variable.
Suppose the Gaussian random variable is of mean $0$ and standard deviation $\sigma$, i.e.,~
$W\sim \mathcal{N}(0,\sigma^2)$.
Then, basically the received value $Y_x$ is also a Gaussian random variable with mean $x$ and
standard deviation $\sigma$, i.e.,~$Y_x\sim \mathcal{N}(X,\sigma^2)$. Note that the random variable $Y$ at the destination (considering all possible $x$) may not be Gaussian, only the conditional distribution at a specific $x$, i.e.,~ $Y \mid X=x$, is the Gaussian $Y_x$.

Suppose the variance of $X$ is $S$ and the density function of $X$ is $f(x)$. We first have
$$
H(Y\mid X=x)= H(Y_x)=\log \sqrt{2\pi {e} \sigma^2},
$$
based on Proposition~\ref{entropy-continuous}. Then,
by definition, we have
$$
H(Y \mid X)=\text{E}_x \{H(Y\mid X=x)\}=\int f(x) \log \sqrt{2\pi {e} \sigma^2} \text{d}x=\log \sqrt{2\pi {e} \sigma^2}.
$$
Hence, the channel capacity per symbol is
$$
\max_X \{H(Y)-H(Y\mid X)\}=\max_X H(Y)-\log \sqrt{2\pi {e} \sigma^2},
$$
which reduces to finding $\max_X H(Y)$.

Note that $Y=X+W$, the sum of two independent random variables, with variances $S$ and $\sigma^2$ respectively. So $Y$ is a random variable with variance $S+\sigma^2$. According to Proposition~\ref{entropy-continuous}, the desired maximum is attained when $Y$ is Gaussian, which makes $X$ the Gaussian variable with variance $S$. Therefore,
$$
\max_X \{H(Y)-H(Y\mid X)\}=\log \sqrt{2\pi {e} (S+\sigma^2)}-\sqrt{2\pi {e} \sigma^2}=\log \sqrt{1+\frac{S}{\sigma^2}}.
$$

What we have just obtained is the capacity per symbol time. Next we ask what is the capacity per second. At this point, if suffices to consider how many symbols can be send in a second.
From Nyquist sampling rate and intersymbol interference criterion, we know that we can send at most $2W$ independent and ISI-free symbols in a second on a channel of bandwidth $W$. 
Namely, if we transmit more than $2W$ symbols in a second, not only will there be unnecessary redundancy, but also this will introduce additional interference besides Gaussian noise. Hence, the channel capacity per second is
$$
2W\cdot \log \sqrt{1+\frac{S}{\sigma^2}} =W \log \left(1+\frac{S}{\sigma^2}\right),
$$
where $\frac{S}{\sigma^2}$ is usually called signal-to-noise ratio (SNR).

\section{Issues Not Usually Addressed}

There are many articles and news claiming that the Shannon capacity limit defined above has been broken.
In fact, these are just kind of advertisements on new technologies with more advanced settings than that of Shannon's original theory,
e.g., multiple-antenna transmitting and receiving technologies (MIMO). Essentially, these technologies are still based on the Shannon capacity, and
they have not broken the Shannon capacity limit at all.

However, here we would like to open up some discussion that seems rarely touched.
There is no problem to model information sources as random processes, i.e.,~sequences.
However, given a channel and a set $\mathbb{A}$ of letters transmittable on the channel,
in order to discuss the capacity of the channel, why are we only allowed to select sequences obeying the same probability distribution as discussed in the last sections? What will happen if the pool of sequences where we are allowed to pick a subset of sequences as signals to transmit on the channel has two sequences in which a same letter in $\mathbb{A}$ may appear with different probabilities? For instance, $\mathbb{A}=\{0,1\}$, and $20\%$ of entries in one sequence are $0$ while $30\%$ of entries in the other sequence are $0$.

\begin{definition}
	Given two probability distributions (random variables) $X_1$ and $X_2$ on $\mathbb{A}=\{x_1, x_2, \ldots, x_n\}$ and a fixed channel (i.e.,~the transit probabilities are fixed),
	if there exists $x_i$ for some $1\leq i \leq n$ such that
	\begin{align}
	\sum_{x_j\in \mathbb{A}} P(X_1=x_j)p(x_i \mid x_j) \neq \sum_{x_j\in \mathbb{A}} P(X_2=x_j)p(x_i \mid x_j),
	\end{align}
then $X_1$ and $X_2$ are called \emph{compatible} (with respect to the channel).
	
\end{definition}

$X_1$ and $X_2$ being compatible implies that the induced random variable at the destination from $X_1$ and $X_2$ are not the same.
In this case, if we transmit any sequence of length $T\rightarrow \infty$ satisfying the distribution $X_1$ and any another sequence of length $T$ satisfying the distribution
$X_2$, the destination should know, by inspecting the number of $x_i$ in the received sequence,
that the transmitted sequence is from the $X_1$-class
or the $X_2$-class. Obviously, if we are allowed to choose sequences from all sequences either satisfying distribution $X_1$
 or $X_2$, we can single out approximately
 $$
 2^{T(H(X_1)-H(X_1\mid Y(X_1))+o(1))}+2^{T(H(X_2)-H(X_2\mid Y(X_2))+o(1))}
 $$
 sequences that can be transmitted on the channel and fully recovered at the destination.

 \begin{definition}
Given $\mathbb{A}=\{x_1, x_2, \ldots, x_n\}$ and a fixed channel,
a set of mutually compatible distributions on $\mathbb{A}$ is called an \emph{admissible set}.
\end{definition}

Then, in theory, the maximal number of sequences of length $T$ ($T\rightarrow \infty$) that are distinguishable at the destination is
$$
\max_F \sum_{X\in F}2^{T(H(X)-H(X|Y)+o(1))},
$$
where $F$ ranges over all admissible sets on $\mathbb{A}$.
As a consequence, it is reasonable to define the channel capacity to be
	\begin{align}\label{newcap}
	\widetilde{C}= \lim_{T\rightarrow \infty}\frac{\log \{\max_F \sum_{X\in F}2^{T(H(X)-H(X|Y)+o(1))}\}}{T}.
	\end{align}

	Intuitively, there is no reason that we cannot have an admissible set containing more than one probability distribution. Thus,
we should potentially have more distinguishable sequences than the number given by the Shannon capacity.
This is exciting, as there is a chance that $\widetilde{C} >C$ whence the Shannon limit is broken.
However, we will argue that this is unfortunately not the case.

\begin{theorem}
 $\widetilde{C} =C$.
\end{theorem}
\proof First, for a sufficient large $T$, two different probability distributions on $\mathbb{A}$
can be characterized by two subsets of sequences of length $T$ on $\mathbb{A}$ where
$x_i$ appears the same number of times in any two sequences in the same subset for any $1\leq i \leq n$ while $x_i$ appears different number of times in any two sequences coming from distinct subset for some $1\leq i \leq n$. Thus, the number of different distributions on $\mathbb{A}$ is equal to the number of integer solutions of the equation $z_1+z_2+\cdots+z_n=T$.
The latter is clearly given by ${T-1 \choose n-1}\approx T^{n-1}$.
Note that for any distribution $X_1$ on $\mathbb{A}$, by construction we have
$$
C \geq H(X_1)-H(X_1 \mid Y(X_1)).
$$
Also note that an admissible set can contain at most $T^{n-1}$ distributions.
Therefore,
$$
2^{T[C+o(1)]} \leq \max_{F}\sum_{X\in F}2^{T[H(X)-H(X|Y(X))+o(1)]} \leq T^{n-1} 2^{T[C+o(1)]}.
$$
Accordingly,
$$
C=\lim_{T\rightarrow \infty} \frac{\log 2^{T[C+o(1)]}}{T} \leq \widetilde{C} \leq \lim_{T\rightarrow \infty} \frac{\log T^{n-1} 2^{T[C+o(1)]}}{T}=C+ (n-1)\lim_{T\rightarrow \infty} \frac{\log T}{T}=C,
$$
and the proof follows. \qed

It is a surprise that there is no gain in terms of capacity although we have more candidate sequences to transmit. {\bf As for the possibility of breaking the Shannon limit, along this line of discussion, it remains to study the case where the source alphabet is infinite or continuous, which is still open.}

\section*{Acknowledgments}
The author thanks Andrei Bura for reading through some early version of the manuscript.

\end{document}